# On a Certain $NP$-Complete Problem

Stepan G. Margaryan

*e-mail: stmargaryan@gmail.com*
*Yerevan, Armenia*

## Abstract:

We intend to create new concepts aimed at finding necessary and sufficient conditions for Boolean satisfiability so that these conditions can be verified in polynomial time. Based on these conditions it will be possible to create an algorithm that determines in polynomial time whether a given Boolean formula represented in conjunctive normal form is satisfiable. The work will consist of three articles. This is the first of a planned series of these articles.

In this article we introduce the concept of special decomposition of a set and the concept of special covering for a set under such a decomposition.

We formulate the decision problem of existence of a special covering for a set under a special decomposition of this set.

In order to determine the complexity class in which this problem is located, we study the relationship between the *sat* $CNF$ problem and the problem of existence of a special covering for a set under a special decomposition of this set.

The article proves that the decidability of the *sat* $CNF$ problem is polynomially reduced to the problem of the existence of a special covering for a set.

It is also proved that the problem of existence of a special covering for a set is polynomially reduced to the decidability of the *sat* $CNF$ problem.

Therefore, the mentioned problems are polynomially equivalent. And then, the problem of existence of a special covering for a set is an $NP$-complete problem.

The conditions for the existence of special coverings for a set under special decompositions of this set will be study in the next articles.

## 1. Special Coverings of Sets

Let $S = \{e_1, e_2, \ldots, e_m\}$ be a nonempty set of $m$ elements, for some natural number $m$.

It is assumed that for the set $S$ and for some natural number $n$ we have $n$ arbitrary ordered pairs of arbitrary subsets of the set $S$.

For some $\alpha \in \{0,1\}$ we denote these ordered pairs by

$$(M_1^{\alpha},\ M_1^{1-\alpha}), (M_2^{\alpha},\ M_2^{1-\alpha}), \ldots, (M_n^{\alpha},\ M_n^{1-\alpha}),$$

where the superscript 1- $\alpha$ means 1 when $\alpha = 0$, and 0, when $\alpha = 1$.

We also denote by $d_nS$ an arbitrarily ordered set of these ordered pairs:

$$d_nS = \{(M_1^{\alpha},\ M_1^{1-\alpha}), (M_2^{\alpha},\ M_2^{1-\alpha}), \ldots, (M_n^{\alpha},\ M_n^{1-\alpha})\}.$$

*Definition* 1.1. The set $d_nS$ will be called a special decomposition of the set $S$, if

(1.1.1) $\forall i \in \{1, \ldots, n\}\ (M_i^{\alpha} \cap M_i^{1-\alpha}) = \emptyset$,

(1.1.2) $\forall i \in \{1, \ldots, n\}\ (M_i^{\alpha} \neq \emptyset$ or $M_i^{1-\alpha} \neq \emptyset)$,

(1.1.3) $\cup_{i=1}^{n}(M_i^{\alpha} \cup M_i^{1-\alpha}) = S$.

The subsets

$$M_1^{\alpha}, M_1^{1-\alpha}, M_2^{\alpha}, M_2^{1-\alpha}, \ldots, M_n^{\alpha},\ M_n^{1-\alpha}$$

of the set $S$ will also be called subsets of the given decomposition.

It is easy to assume, that the same subsets of the set can form different special decompositions and also these subsets may not allow any special decompositions.

*Definition* 1.2. Let the set $d_nS$ be a special decomposition of the set $S$. An ordered set

$$c_nS = \{M_1^{\alpha_1}, M_2^{\alpha_2}, \ldots, M_n^{\alpha_n}\},\ (\alpha_i \in \{0,1\}),$$

will be called a special covering for the set $S$ under the decomposition $d_nS$, if

$$\cup_{i=1}^{n} M_i^{\alpha_i} = S.$$

It follows from the Definition 1.2, that for any $i \in \{1, \ldots, n\}$ the subsets $M_i^{\alpha}$ and $M_i^{1-\alpha}$ cannot simultaneously belong to the covering, but one of them exactly belongs.

*Proposition* 1.3. For some $\alpha_1, \ldots, \alpha_n$ the set

$$c_nS = \{M_1^{\alpha_1}, M_2^{\alpha_2}, \ldots, M_n^{\alpha_n}\}$$

is a special covering for the set $S$ under the special decomposition $d_nS$ if and only if for any $i \in \{1, \ldots, n\}$ and for any $e \in M_i^{1-\alpha_i}$, there exists a subset $M_j^{\alpha_j}$ such, that

$$(M_j^{\alpha_j} \in c_nS)\ \&\ (j \neq i)\ \&\ (e \in M_j^{\alpha_j}).$$

*Proof.* Since the set $c_nS$ is a special covering for the set $S$, then obviously,

$$M_i^{1-\alpha_i} \subseteq M_1^{\alpha_1} \cup \ldots \cup M_{i-1}^{\alpha_{i-1}} \cup M_{i+1}^{\alpha_{i+1}} \cup \ldots \cup M_n^{\alpha_n}.$$

This means that for any element $e \in M_i^{1-\alpha_i}$ there exists a subset $M_j^{\alpha_j} \in cS\ (j \neq i)$ such, that $e \in M_j^{\alpha_j}$. Obviously, the opposite is also true. $\nabla$

(By the symbol $\nabla$ we mark the end of the proof).

*Proposition* 1.4. Let for some $\alpha_1, \alpha_2, \ldots, \alpha_n$, the set

$$c_nS = \{M_1^{\alpha_1}, M_2^{\alpha_2}, \ldots, M_n^{\alpha_n}\}$$

be a special covering for the set $S$ under the special decomposition $d_nS$, where $\alpha_i \in \{0,1\}$.

If $M_i^{\alpha} \nsubseteq \bigcup_{j\neq i}(M_j^{\alpha_j} \cup M_j^{1-\alpha_j})$ for some $\alpha \in \{0,1\}$, then $M_i^{\alpha} \in c_nS$.

*Proof.* Suppose that $M_i^{\alpha} \notin c_nS$. It means that $M_i^{1-\alpha} \in c_nS$.
Since $M_i^{\alpha} \not\subset \bigcup_{j\neq i}(M_j^{\alpha_j} \cup M_j^{1-\alpha_j})$, then there exists an element $e \in M_i^{\alpha}$, such that

$$\forall\ j \neq i\ \ (\, e \notin M_j^{\alpha_j}\,)\ \&\ (\, e \notin M_j^{1-\alpha_j}\,).$$

On the other hand, since $M_i^{\alpha} \cap M_i^{1-\alpha} = \emptyset$, then it follows from $e \in M_i^{\alpha}$ that $e \notin M_i^{1-\alpha}$.

So, $c_nS$ cannot be a special covering for the set $S$. And this is a contradiction. ∇

*Corollary* 1.4.1. If under some special decomposition $d_nS$ of the set $S$ there is an ordered pair $(M_i^{\alpha},\ M_i^{1-\alpha}) \in d_nS$ such, that

$$M_i^{\alpha} \nsubseteq \bigcup_{j\neq i}(M_j^{\alpha_j} \cup M_j^{1-\alpha_j}) \text{ and } M_i^{1-\alpha} \nsubseteq \bigcup_{j\neq i}(M_j^{\alpha_j} \cup M_j^{1-\alpha_j}),$$

then there is no special covering of the set $S$ under the decomposition $d_nS$.

*Proof.* If under the mentioned conditions there is a special covering, then it follows from the Proposition 1.4 that both subsets $M_i^{\alpha}$ and $M_i^{1-\alpha}$ should be included in it. But this is contrary to the Definition 1.2. ∇

Let for some natural number $n$

$$d_nS = \{(M_1^{\alpha},\ M_1^{1-\alpha}), \ldots, (M_i^{\alpha},\ M_i^{1-\alpha}), \ldots, (M_n^{\alpha},\ M_n^{1-\alpha})\}$$

be an ordered set of arbitrary ordered pairs of subsets of the set $S$.

Any permutation $(i_1, i_2, \ldots, i_n)$ of elements of the set $d_nS$, when the orders of pairs do not change, we will call $P$-transformation of $d_nS$, where $i_j$ is the number of the pair that moves to the $j$-th place.

We denote by $(i_1, i_2, \ldots, i_n)P(d_nS)$ the resulting ordered set obtained by means of $P$-transformation of the set $d_nS$. If there is no need to mark the numbers of ordered pairs, then the $P$-transformation of the set $d_nS$ will be denoted by $P(d_nS)$:

$$P(d_nS) = \{(M_{i_1}^{\alpha},\ M_{i_1}^{1-\alpha}), \ldots, (M_{i_n}^{\alpha},\ M_{i_n}^{1-\alpha})\},$$

where

$$(M_{i_k}^{\alpha}, M_{i_k}^{1-\alpha}) = \begin{cases} (M_k^{\alpha},\ M_k^{1-\alpha}), & \text{if the } k-\text{th element of } d_nS \text{ is not moved,} \\ (M_j^{\alpha},\ M_j^{1-\alpha}), & \text{if the } j-\text{th element moves to the } k-\text{th place of the set } d_nS \end{cases}$$

For any $k$ $(1 \leq k \leq n)$, the permutation of the components of ordered pairs

$$(M_{i_1}^{\alpha}, M_{i_1}^{1-\alpha}), \ldots, (M_{i_k}^{\alpha},\ M_{i_k}^{1-\alpha})$$

of the set $d_nS$, when the orders of the elements of $d_nS$ do not change, will be called an $I$-transformation of the set $d_nS$.

The ordered set obtained as a result of $I$- transformation will be denoted as

$$(i_1,i_2,\ldots,i_k)I(d_nS).$$

If it is not necessary to mark the numbers of the pairs participating in the transformation, then *I*-transformation of the set $d_nS$ will be denoted by $I(d_nS)$:

$$I(d_nS) = \{(M_1^{\alpha_1}, M_1^{1-\alpha_1}), \ldots, (M_i^{\alpha_i}, M_i^{1-\alpha_i}), \ldots, (M_n^{\alpha_n}, M_n^{1-\alpha_n})\}$$

where $\alpha_i \in \{0,1\}$ and

$$\left(M_i^{\alpha_i}, M_i^{1-\alpha_i}\right) = \begin{cases} (M_i^{\alpha}, M_i^{1-\alpha}), \text{if the components of the } i-\text{th pair are not displaced,} \\ (M_i^{1-\alpha}, M_i^{\alpha}), \text{if the components of the } i-\text{th pair are displaced.} \end{cases}$$

Obviously, the transition from the set $P(d_nS)$ to the set $d_nS$ is a $P$-transformation, and the transition from the set $I(d_nS)$ to the set $d_nS$ is an $I$-transformation.

*Lemma* 1.5. If $d_nS$ is an ordered set of ordered pairs of subsets of the set $S$ then for any $P$ -transformation $P(d_nS)$, the following is true:

(i) the set $d_nS$ is a special decomposition of the set $S$ if and only if $P(d_nS)$ is a special decomposition of the set $S$.

(ii) If the set $d_nS$ is a special decomposition of the set $S$, then there exists a special covering for the set $S$ under the decomposition $d_nS$ if and only if it exists under the decomposition $P(d_nS)$.

*Proof.* i) Obviously, during the transition from the set $d_nS$ to the set $P(d_nS)$ and the transition from $P(d_nS)$ to $d_nS$ the contents of the subsets of the decomposition do not change, but only the orders of the elements of $d_nS$ changes. This means, that the conditions of special covering are not violated. Therefore, the points i) and ii) are true. ∇

*Lemma* 1.6. If $d_nS$ is an ordered set of ordered pairs of subsets of the set $S$ then for any *I*-transformation $I(d_nS)$, the following is true:

i) $d_nS$ is a special decomposition of the set $S$ if and only if $I(d_nS)$ is a special decomposition of the set $S$.

(ii) If $d_nS$ is a special decomposition of the set $S$, then there exists a special covering for the set $S$ under the decomposition $d_nS$ if and only if it exists under the decomposition $I(d_nS)$.

*Proof.* i) During the transition from the set $d_nS$ to the set $I(d_nS)$ and the transition from $I(d_nS)$ to $d_nS$, the contents of the subsets of decomposition do not change. Only the orders of the components of some ordered pairs change.

Therefore, the sets $d_nS$ and $I(d_nS)$ are either at the same time special decompositions of the set $S$, or at the same time they are not such decompositions. Also, it is obvious that if under decomposition $d_nS$, the set

$$c_nS=\{M_1^{\alpha_1},\ldots,M_i^{\alpha_i},\ldots,M_n^{\alpha_n}\} \quad (\alpha_i \in \{0,1\})$$

is a special covering for the set $S$, then it will also be a special covering for the set $S$ under decomposition $I(d_nS)$ and vice versa. So, the points i) and ii) are true. ∇

According to Lemmas 1.5 and 1.6, for any special decomposition of the set $S$, any $P$-transformation or $I$-transformation preserve the possibility of being a special decomposition of the set $S$ and having a special covering for $S$ under such a decomposition.

For technical convenience, for any $\alpha \in \{0,1\}$ we denote:

$M^{\alpha} = \bigcup_{i=1}^{n} M_i^{\alpha}$,

$sM^{\alpha} = \{M_1^{\alpha}, M_2^{\alpha}, \ldots, M_n^{\alpha}\}$,

$(i_1, \ldots, i_k)sM^{\alpha}$ is the set obtained by replacing the subsets $M_{i_1}^{1-\alpha}, \ldots, M_{i_k}^{1-\alpha}$ of the set $sM^{\alpha}$ by the subsets $M_{i_1}^{\alpha}, \ldots, M_{i_k}^{\alpha}$, respectively.

Note that $sM^{\alpha}$ and $(i_1, \ldots, i_k)sM^{\alpha}$ we consider as ordered sets.

*Definition* 1.7. i) The set $sM^{\alpha}$ will be called a set of $\alpha$-components of ordered pairs of the decomposition.

ii) For any $\{i_1, \ldots, i_k\} \subseteq \{1, \ldots, n\}$ the set $(i_1, \ldots, i_k)sM^{\alpha}$ is called a set of $\alpha$-components of the ordered pairs of decomposition $(i_1, \ldots, i_k)I(d_nS)$.

iii) For any decomposition, the set of $\alpha$-components of ordered pairs will also be called a set of subsets of the $\alpha$-domain.

iv) If the set of subsets of the $\alpha$-domain is a special covering for the set $S$, then such a covering will be called a special $M^{\alpha}$-covering or briefly $M^{\alpha}$-covering for the set $S$.

*Lemma* 1.8 Let the set

$$d_nS = \{(M_1^{\alpha}, M_1^{1-\alpha}), \ldots, (M_i^{\alpha}, M_i^{1-\alpha}), \ldots, (M_n^{\alpha}, M_n^{1-\alpha})\}$$

be a special decomposition of the set $S$.

Then, there exists a special covering for the set $S$ under the special decomposition $d_nS$ if and only if for some $\alpha \in \{0,1\}$, there exists an $M^{\alpha}$-covering for the set $S$ under some special decomposition $I(d_nS)$.

*Proof.* Obviously, for any $\alpha \in \{0,1\}$, the procedure for forming the $\alpha$-domain does not violate the Definition 1.2 of a special covering. Therefore, an $M^{\alpha}$-covering is also a special covering for the set $S$.

Now suppose that there is a special covering for the set $S$. Let it be the set

$$c_nS = \{M_1^{\alpha_1}, M_2^{\alpha_2}, \ldots, M_n^{\alpha_n}\}.$$

If $M_i^{\alpha_i} \in sM^{\alpha}$ for any $i \in \{1, \ldots, n\}$, then $c_nS$ is also an $M^{\alpha}$-covering.

If $M_{j_1}^{1-\alpha}, \ldots, M_{j_l}^{1-\alpha}$ are subsets such, that

$$(\{M_{j_1}^{1-\alpha}, \ldots, M_{j_l}^{1-\alpha}\} \subseteq sM^{1-\alpha}) \,\&\, (\{M_{j_1}^{1-\alpha}, \ldots, M_{j_l}^{1-\alpha}\} \subseteq c_nS),$$

then applying $I$-transformation with respect to the ordered pairs

$$(M_{j_1}^{\alpha}, M_{j_1}^{1-\alpha}), \ldots, (M_{j_l}^{\alpha}, M_{j_l}^{1-\alpha}),$$

according to Lemma 1.6, we obtain that $c_nS$ is also an $M^{\alpha}$-covering for the set $S$. ∇

## 2. The $NP$-Completeness of the Problem of a Special Covering

Consider Boolean functions.

Let a Boolean function $f(x_1, \ldots, x_n)$ of $n$ variables be given in conjunctive normal form ($CNF$) with $m$ clauses. With a certain natural numbering of clauses, we denote by $c_i$ the $i$-th clause of the formula. That is

$$c_i = x_{j_1}^{\alpha_1} \vee \ldots \vee x_{j_k}^{\alpha_k}, \text{ where } \{ j_1, \ldots, j_k\} \subseteq \{1, \ldots, n\},\ k \in \{1, \ldots, n\},\ \alpha_j \in \{0,1\},$$

$$x_j^0 = \neg x_j,\quad x_j^1 = x_j,\quad j \in \{1, \ldots, n\}.$$

$$f(x_1, \ldots, x_n) = \wedge_{i=1}^{m} c_i.$$

For simplicity, we assume that no variable and its negation are included in any clause simultaneously. Obviously, it does not limit the set of functions being considered.

The $sat\,CNF$ problem is the problem of determining if given function $f$ is satisfiable, that is if there exist $\sigma_1, \ldots, \sigma_n$, such that $\sigma_j \in \{0, 1\}$ and

$$f(\sigma_1, \sigma_2, \ldots, \sigma_n) = 1$$

Our goal is to reduce the decidability of the $sat\,CNF$ problem to the decidability of the problem of finding a special covering for a set under the special decomposition of the set.

To discuss the possibilities of reducibility, we need to define specific formal languages over some alphabets. For convenience we consider the alphabet

$$\Sigma = \{0, 1, x^0, x^1, \star, ▫, e, \varepsilon\}$$

and other alphabets included in $\Sigma$.

We use the symbols 0, 1, $x^0$, $x^1$, to form different literals. We consider literals in following meanings:

$$x_i^0 = x^0 b(i) = \neg x_i \quad \text{and} \quad x_i^1 = x^1 b(i) = x_i,$$

where $b(i)$ is binary code of number $i$, in any one to one binary encoding. We will identify $x_i$ and $x_i^1$.

The expression $b(i)$ will be called the index of the literal.

If a function of $n$ variables is represented in $CNF$, then with this designation, $n$ will also be the number of literals with pairwise different indices included in the clauses of the function.

Similarly, we will use the symbols $e$, 0, 1 to form expressions such as $e_i = eb(i)$, for designation elements of sets, where $b(i)$ is the binary code of number $i$.

However, for convenience, when assigning literals and elements of the sets, we will often use subscripts.

The symbols '⋆' and '▫' are separating symbols. And '$\varepsilon$' is a special symbol, which will correspond to empty string.

For any natural numbers $n$ and $m$ we define the languages $CLAUSE(n)$ and $CNF(n, m)$ over the alphabet $\Sigma_1 = \{0, 1, x^0, x^1, \star\} \subseteq \Sigma$ as follows:

$$CLAUSE(n) = \{\langle x_{i_1}^{\alpha_1}\, x_{i_2}^{\alpha_2} \ldots x_{i_k}^{\alpha_k} \rangle \,/\, \{x_{i_1}^{\alpha_1}, x_{i_2}^{\alpha_2}, \ldots, x_{i_k}^{\alpha_k}\} \text{ is a set of literals, } (1 \leq k \leq n) \ \& \ \& \ (1 \leq i_j \leq n) \ \& \ (\alpha_i \in \{0,1\}) \ \& \ (i_r \neq i_l, \text{ if } r \neq l)\}.$$

It is important to note, that the language $CLAUSE(n)$ does not contain an empty string. Also, it does not contain a string that includes any variable and its negation simultaneously.

We will say that the string

$$\langle\, x_{i_1}^{\alpha_1}\, x_{i_2}^{\alpha_2} \;.\;.\;.\; x_{i_k}^{\alpha_k} \,\rangle \text{ corresponds to the set } \{x_{i_1}^{\alpha_1}, x_{i_2}^{\alpha_2}, \ldots, x_{i_k}^{\alpha_k}\}$$

as well as we will say that the string is formed by the clause.

Let's denote by $c_{j_i}$ with different indices, the strings included in $CLAUSE(n)$.

$$c_{j_i} = \langle\, x_{i_1}^{\alpha_1}\, x_{i_2}^{\alpha_2} \;.\;.\;.\; x_{i_k}^{\alpha_k} \,\rangle.$$

With such designations we define the language $CNF(n,m)$ as follows:

$$CNF(n,m) = \{\langle\, c_{j_1} \star c_{j_2} \star \ldots \star c_{j_m}\rangle \,/\, \forall\, i \in \{1, \ldots, m\},\; c_{j_i} \in CLAUSE(n) \text{ and } n \text{ is the number of literals with pairwise different indices in clauses}\}.$$

*Remark* 2.1. It is easy to prove, that for any natural numbers $n$ and $m$,

$$\langle\, c_{j_1} \star c_{j_2} \star \ldots \star c_{j_m}\rangle \in CNF(n,m)$$

if and only if the clauses corresponding to the strings $c_{j_1}, c_{j_2}, \ldots, c_{j_m}$, form a certain function $f$ represented in $CNF$ with $n$ variables and with $m$ clauses.

We denote by $N(f)$ the number of literals and symbols '$\star$' of the string

$$\langle\, c_{j_1} \star c_{j_2} \star \ldots \star c_{j_m}\rangle$$

$N(f)$ will also be called the length of the input data for the $CNF$ of the function $f$.

Let's consider the alphabet $\Sigma_2 = \{0, 1, \star, \;▫\;, \; e, \varepsilon\} \subseteq \Sigma$.

Using the symbols $e$, 0 and 1, we compose expressions over the $\Sigma_2$ to denote elements of an arbitrary set:

$$S = \{e_{j_1},\; e_{j_2},\; \ldots, e_{j_m}\},$$

where $j_1,\; j_2, \ldots, j_m$ are different binary expressions.

Since the empty subset of the set plays a certain important role in the special decomposition, we need a symbol for forming the strings corresponding to the pairs of subsets with an empty component. As such a symbol we will consider the symbol $\varepsilon \in \Sigma$.

That is, we will use the string $\varepsilon$ to form strings corresponding to pairs of subsets such as

$$(M, \emptyset) \text{ or } (\emptyset, M)$$

for some nonempty subset $M$.

For a natural number $m$, we define the languages $SAB(m)$, $SAB^{\varepsilon}(m)$ and $PAIR(m)$ over the alphabet $\Sigma_2$ as follows:

$$SAB(m) = \{\langle\, e_{j_1}\, e_{j_2} \;\ldots\; e_{j_k} \,\rangle / \,(1 \le k \le m) \,\&\, (j_p \ne j_q, \text{ if } p \ne q)\}.$$

$$SAB^{\varepsilon}(m) = \{\varepsilon\} \cup SAB(m).$$

Note that the language $SAB(m)$ does not contain an empty string.

We say that the string $M = \langle\, e_{j_1}\, e_{j_2} \;\ldots\; e_{j_k} \,\rangle$ corresponds to the set $\{M\} = \{\, e_{j_1}, e_{j_2}, \ldots, e_{j_k}\}$ or the string $M$ is formed by the set $\{M\}$.

It is obvious that an element is included in the string $M$ of the language $SAB(m)$, if and only if it belongs to the set $\{M\}$, and $\{M\}$ forms the string $M$.

For any set $N$, we denote by $\langle N\rangle$ the string that is formed by the set $N$. That is for a set $N$ of at most $m$ elements $\langle N\rangle \in SAB(m)$.

With such designation we define the following:

$$PAIR(m) = \{\langle M \circ N\rangle \,/\, (M \in SAB^{\varepsilon}(m))\ \&\ (N \in SAB^{\varepsilon}(m))\ \&\ (\{M\} \cap \{N\} = \emptyset)\ \&\ \&\ ((M \neq \varepsilon) \text{ or } (N \neq \varepsilon))\}.$$

For any strings $M \in SAB^{\varepsilon}(m)$ and $N \in SAB^{\varepsilon}(m)$, the string $\langle M \circ N\rangle$ included in the language PAIR$(m)$, is an ordered pair of $M$ and $N$.

It is important to note, that for any natural number $m$, holds

$$\langle M \circ \varepsilon\rangle \in PAIR(m) \quad \text{and} \quad \langle \varepsilon \circ N\rangle \in PAIR(m)$$

only with $\{M\} \neq \emptyset$ and $\{N\} \neq \emptyset$.

For any natural numbers $n$ and $m$, we define the language $d(n, m)$ over the alphabet $\Sigma_2$ as follows:

$$d(n, m) = \{\langle M_1^{\alpha} \circ M_1^{1-\alpha} \star \ldots, \star M_i^{\alpha} \circ M_i^{1-\alpha} \star \ldots \star M_n^{\alpha} \circ M_n^{1-\alpha}\rangle /$$

$m$ is the number of elements with pairwise different indices in all subsets $M_i^{\alpha_i}$ and $\forall i \in \{1, \ldots, n\} \langle M_i^{\alpha} \circ M_i^{1-\alpha}\rangle \in PAIR(m)\}$.

For any string

$$\langle dS\rangle = \langle M_1^{\alpha} \circ M_1^{1-\alpha} \star \ldots \star M_i^{\alpha} \circ M_i^{1-\alpha} \star \ldots \star M_n^{\alpha} \circ M_n^{1-\alpha}\rangle \in d(n, m),$$

we call the sets $\{M_i^{\alpha_i}\}$ (for $\alpha_i \in \{0, 1\}$ and $i \in \{1, \ldots, n\}$), the subsets corresponding to $\langle dS\rangle$.

*Remark* 2.2. It is obvious, that for any natural numbers $n$ and $m$,

$$\langle M_1^{\alpha} \circ M_1^{1-\alpha} \star \ldots \star M_i^{\alpha} \circ M_i^{1-\alpha} \star \ldots \star M_n^{\alpha} \circ M_n^{1-\alpha}\rangle \in d(n, m)$$

if and only if $m$ is the number of elements of the set

$$\bigcup_{i=1}^{n} (\{M_i^{\alpha}\} \cup \{M_i^{1-\alpha}\})$$

and the ordered set of ordered pairs

$$\{(\{M_1^{\alpha}\}, \{M_1^{1-\alpha}\}), \ldots, (\{M_i^{\alpha}\}, \{M_i^{1-\alpha}\}), \ldots, (\{M_n^{\alpha}\}, \{M_n^{1-\alpha}\})\},$$

is a special decomposition of the set of $m$ elements, denoted by

$$S = \bigcup_{i=1}^{n} (\{M_i^{\alpha}\} \cup \{M_i^{1-\alpha}\}),$$

where $\langle M_i^{\alpha}\rangle$ corresponds to the set $\{M_i^{\alpha}\}$ for $i \in \{1, \ldots, n\}$.

Sometimes, if it does not lead to an ambiguity, we will use the notation $M$ instead of $\{M\}$ for the set corresponding to the string $\langle M\rangle$.

Let the function $f(x_1, \ldots, x_n)$ be represented in $CNF$ with $m$ clauses, $n$ is the number of literals with pairwise different indices included in the clauses of $f$, and $S$ be a nonempty set of $m$ elements.

We match the elements of the set $S$ to the clauses of the function $f(x_1, \ldots, x_n)$ in some sequential enumeration of clauses so that the element $e_j = eb(j)$ corresponds to the $j$-th clause of the function $f(x_1, \ldots, x_n)$.

We will identify the clause and its corresponding element $eb(i)$, which we will call the number of the clause, or the code of the clause. So, for brevity, we denote the set of clauses of the function $f(x_1, \ldots, x_n)$ as

$$S(f) = \{e_1,\ e_2, \ldots, e_m\}.$$

We will use the notation $x_i^{\alpha_i} \in e_j$, considering it as the inclusion of the literal $x_i^{\alpha_i}$ in the clause corresponding to $e_j$.

For some $\alpha \in \{0, 1\}$ and for each $i \in \{1, \ldots, n\}$ we form the following sets:

$$fM_i^{\alpha} = \{e_j\ /\ e_j \in S(f)\ \text{ and }\ x_i^{\alpha} \in e_j,\ \ (j \in \{1, \ldots, m\})\},$$
$$fM_i^{1-\alpha} = \{e_j\ /\ e_j \in S(f)\ \text{ and }\ x_i^{1-\alpha} \in e_j,\ \ (j \in \{1, \ldots, m\})\}.$$

Obviously, for any $\alpha \in \{0, 1\}$ and $i \in \{1, \ldots, n\}$, the sets $fM_i^{\alpha}$ and $fM_i^{1-\alpha}$ are subsets of the set $S(f)$. At the same time, if $x_i^{\alpha} \notin e_j$ for some $\alpha \in \{0, 1\}$ and for all $j \in \{1, \ldots, m\}$, then $fM_i^{\alpha} = \emptyset$. Let's denote

$$f_n = \langle\ e_1 \star\ e_2 \star\ \ldots\ \star e_m\ \rangle,$$

where $\langle e_j \rangle \in CLAUSE(n)$ is the string formed by clause $e_j$ of function $f$.

That is, we will consider $f_n$ as a string corresponding to the function $f$, and $f$ will be considered as a function corresponding to the string $f_n$.

*Proposition* 2.3. If $f_n \in CNF(n, m)$ for some natural numbers $n$ and $m$, then

i) $\forall i \in \{1, \ldots, n\}\ \langle fM_i^{\alpha} \circ fM_i^{1-\alpha} \rangle \in PAIR(m)$.

ii) $\bigcup_{i=1}^{n} (fM_i^{\alpha} \cup fM_i^{1-\alpha}) = S(f)$.

*Proof.* i) We will prove, that for any $i \in \{1, \ldots, n\}$,

i.1) $\langle fM_i^{\alpha} \rangle \in SAB^{\varepsilon}(m)$ and $\langle fM_i^{1-\alpha} \rangle \in SAB^{\varepsilon}(m)$,

i.2) $fM_i^{\alpha} \cap fM_i^{1-\alpha} = \emptyset$,

i.3) $fM_i^{\alpha} \neq \emptyset$ or $fM_i^{1-\alpha}) \neq \emptyset$.

i.1) Since for any $i \in \{1, \ldots, n\}$, the sets $fM_i^{\alpha}$ and $fM_i^{1-\alpha}$ are subsets of the set $S(f)$, then the point i.1) holds.

i.2) If $fM_i^{\alpha} \cap fM_i^{1-\alpha} \neq \emptyset$ for some pair $(fM_i^{\alpha},\ fM_i^{1-\alpha})$, then there exists a clause, with certain number $e_{j_l}$, which contains $x_i^{\alpha}$ and $x_i^{1-\alpha}$ simultaneously. This contradicts the condition of the language $CLAUSE$ $(n)$.

i.3) Suppose there exists an index $i \in \{1, \ldots, n\}$ such, that $fM_i^{\alpha} = \emptyset$ and $fM_i^{1-\alpha} = \emptyset$.

It means the literals $x_i^{\alpha}$ and $x_i^{1-\alpha}$ do not belong to any of clauses. But it contradicts to condition of language $CNF(n, m)$, the procedure of forming subsets $fM_i^{\alpha}$ and $fM_i^{1-\alpha}$ as well as the condition of the language $CLAUSE$ $(n)$.

ii) If $e_j \in \bigcup_{i=1}^{n}(fM_i^{\alpha} \cup fM_i^{1-\alpha})$, then, there is an index $i \in \{1, \ldots, n\}$ such that

$$e_j \in fM_i^{\alpha} \text{ or } e_j \in fM_i^{1-\alpha} \text{ and therefore, } e_j \in S(f).$$

If $e_j \in S(f)$ then $e_j$ contains some literals. It means there is an index $i \in \{1, \ldots, n\}$ such, that $x_i^{\alpha}$ is found in clause $e_j$ or $x_i^{1-\alpha}$ is found in clause $e_j$, which means, that

$$e_j \in fM_i^{\alpha} \cup fM_i^{1-\alpha} \quad \text{and} \quad e_j \in \bigcup_{i=1}^{n}(fM_i^{\alpha} \cup fM_i^{1-\alpha}).$$

Hence

$$\bigcup_{i=1}^{n}(fM_i^{\alpha} \cup fM_i^{1-\alpha}) = S(f). \nabla$$

*Lemma* 2.4. If for any natural numbers $n$ and $m$,

$$f_n = \langle\, e_1 \star\ e_2 \star\ \ldots\ \star e_m \,\rangle \in CNF(n, m),$$

then for any $\alpha \in \{0, 1\}$, the string

$$\langle\, fM_1^{\alpha} \,\square\, fM_1^{1-\alpha} \star fM_2^{\alpha} \,\square\, fM_2^{1-\alpha} \star \ldots \star fM_n^{\alpha} \,\square\, fM_n^{1-\alpha} \,\rangle$$

is formed in polynomial time with respect to the length of string $f_n$.

*Proof.* We describe the general features of the workflow of a particular Turing machine ($TM$), with the alphabet $\Sigma$, which outputs the string

$$\langle\, fM_1^{\alpha} \,\square\, fM_1^{1-\alpha} \star fM_2^{\alpha} \,\square\, fM_2^{1-\alpha} \star \ldots \star fM_n^{\alpha} \,\square\, fM_n^{1-\alpha} \,\rangle$$

in polynomial time, when input the string $f_n$.

It is convenient to consider a Turing machine denoted by $T_1$, with input tape, output tape and with two working tapes. The $TM$ operation procedure consists of following stages.

The input tape receives a string of clauses.

On the first stage of operation, $T_1$ records the codes of clauses on the first working tape in the same order in which the clauses are on the input tape. We do this in order to easily remember the code of the clause being considered.

To form the current strings $fM_i^{\alpha}$ and $fM_i^{1-\alpha}$, $T_1$ uses the second working tape, successively recording the pairs $(i, \alpha)$ and $(i, 1\text{-}\alpha)$ on it for $i \in \{1, \ldots, n\}$.

We say that $T_1$ scans a code or a pair of indices, meaning that it scans the symbols that form them.

After recording the numbers of clauses, $T_1$ starts working, recording $(1, \alpha)$ on the second working tape and $\varepsilon$ on output tape.

Each action of $T_1$ is determined by configuration of four records on different tapes - scanned at current moment. We denote it $(R_1,\ R_2,\ R_3,\ R_4)$, where

$R_1$ is current scanned record on input tape,

$R_2$ is current scanned record on first working tape,

$R_3$ is current scanned record of pair $(i, \alpha)$ on the second working tape, where $i \in \{1, \ldots, n\}$, $\alpha \in \{0,1\}$,

$R_4$ is current scanned record on output tape.

The first configuration is $(\, x_j^{\alpha},\ e_1,\ (1, \alpha), \varepsilon)$, where $x_j^{\alpha}$ is the first symbol of the string $f_n$.

At the beginning of the work, $T_1$ scans these records on the tapes.

To form $fM_i^{\alpha}$ and $fM_i^{1-\alpha}$, $T_1$ performs actions depending on configuration:

with configuration $(x_i^{\alpha},\ e_j,\ (i,\alpha), \varepsilon)$,

- records $e_j$ instead of $\varepsilon$, on the output tape,
- scans the symbol $e_j$ on the output tape,
- scans the symbol following of $x_i^{\alpha}$ on the input tape,

with configuration $(x_i^{\alpha},\ e_j,\ (i,\alpha), e_k)$,

- adds $e_j$ to the output tape and scans it,
- scans the symbol following of $x_i^{\alpha}$ on the input tape,

with configuration $(x_i^{\alpha},\ e_j,\ (s,\alpha), \varepsilon)$ or $(x_i^{\alpha},\ e_j,\ (s,\alpha), e_k)$, where $i \neq s$,

- scans the symbol following of $x_i^{\alpha}$ on the input tape,

with configuration $(\star,\ e_j,\ (i,\alpha), e_k)$,

- scans the symbol $e_{j+1}$ on the first working tape
- scans the symbol following of the symbol '$\star$' on the input tape,

with configuration (*end of input data*, $e_j$, $(i,\alpha), e_k)$,

- adds the symbol '▫' after $e_k$ on the output tape,
- records the pair $(i$, 1-$\alpha)$ on the next place following the pair $(i,\alpha)$ and scans it,
- records the symbol $\varepsilon$ after the symbol '▫' on the output tape and scans it,
- scans the first symbol on input tape,

with configuration $(x_i^{\alpha},\ e_j,\ (s, 1\text{-}\alpha), \varepsilon)$ or $(x_i^{\alpha},\ e_j,\ (s, 1\text{-}\alpha), e_k)$,

- scans the symbol following of $x_i^{\alpha}$ on the input tape,

with configuration (*end of input data*, $e_j$, $(i, 1\text{-}\alpha), e_k)$ and for $i < n$,

- adds the symbol '$\star$' on the next place following the $e_k$ on output tape,
- records the symbol $\varepsilon$ on the next place of '$\star$' on the output tape and scans it,
- records the pair $(i+1, \alpha)$ on the next place following the pair $(i, 1\text{-}\alpha)$ on the second working tape and scans it,
- scans the first symbol on input tape,

with (*end of input data*, $e_j$, $(n, 1\text{-}\alpha), e_k)$, the Turing's machine $T_1$ stops.

It is easy to see that $T_1$ outputs the string

$$\langle\, fM_1^{\alpha} ▫ fM_1^{1-\alpha} \star fM_2^{\alpha} ▫ fM_2^{1-\alpha} \star \ldots \star fM_n^{\alpha} ▫ fM_n^{1-\alpha}\,\rangle.$$

Also, the number of actions performed by $T_1$ to complete the described procedure is polynomial with respect to the length of input data of the string $f_n$, denoted by $N(f)$.

Obviously, the number of mentioned operations does not exceed the number

$$c \times (N(f))^2$$

for certain constant $c$. ∇

For some $\alpha \in \{0, 1\}$ we denote by $dS(f)$ the following set:

$$dS(f) = \{(fM_1^{\alpha}, fM_1^{1-\alpha}), (fM_2^{\alpha}, fM_2^{1-\alpha}), \ldots, (fM_n^{\alpha}, fM_n^{1-\alpha})\}.$$

*Remark* 2.5. In fact, Lemma 2.4 enables us, based on some function $f$ as an input data, to obtain the set $dS(f)$ in polynomial time with respect to the length of mentioned input data.

Let's denote by $T_1(f_n)$ the string formed as a result of $T_1$'s work on input $f_n$:

$$T_1(f_n) = \langle fM_1^{\alpha} \,\square\, fM_1^{1-\alpha} \star fM_2^{\alpha} \,\square\, fM_2^{1-\alpha} \star \ldots \star fM_n^{\alpha} \,\square\, fM_n^{1-\alpha} \rangle.$$

*Lemma* 2.6. For any natural numbers $n$ and $m$,

$$f_n \in CNF(n, m) \quad \text{if and only if} \quad T_1(f_n) \in d(n, m).$$

*Proof.* Obviously, if $f_n \in CNF(n, m)$ for some natural numbers $n$ and $m$, then, according to Proposition 2.3 i),

$$T_1(f_n) = \langle fM_1^{\alpha} \,\square\, fM_1^{1-\alpha} \star fM_2^{\alpha} \,\square\, fM_2^{1-\alpha} \star \ldots \star fM_n^{\alpha} \,\square\, fM_n^{1-\alpha} \rangle \in d(n, m).$$

Now, let $T_1(f_n) \in d(n, m)$ for some natural numbers $n$ and $m$. It means that

i) $\forall i \in \{1, \ldots, n\}$ ( $fM_i^{\alpha} \neq \emptyset$ or $fM_i^{1-\alpha} \neq \emptyset$),

ii) $\forall i \in \{1, \ldots, n\}$ ( $fM_i^{\alpha} \cap fM_i^{1-\alpha} = \emptyset$).

The point i) means, that for each $i \in \{1, \ldots, n\}$, there is a clause $e_j$ such, that $e_j \in fM_i^{\alpha}$ or there is a clause $e_k$ such, that $e_k \in fM_i^{1-\alpha}$.

The point ii) means, that for any $i \in \{1, \ldots, n\}$ there is no clause included in both subsets $fM_i^{\alpha}$ and $fM_i^{1-\alpha}$ at the same time.

In fact, we get the following:

- for each $i \in \{1, \ldots, n\}$ there is a clause including the literal $x_i^{\alpha}$ or there is a clause including the literal $x_i^{1-\alpha}$,

- there is no clause including the literals $x_i^{\alpha}$ and $x_i^{1-\alpha}$ simultaneously.

It means that the number of literals with pairwise different indexes is equal to $n$. Meanwhile, each clause $e_j$, included in some set $fM_i^{\alpha_i}$ cannot be empty. Therefore

$$f_n \in CNF(n, m). \ \nabla$$

If under the special decomposition $dS(f)$, there exists a special covering for the set $S(f)$, then we will denote such a covering by $c_nS(f)$:

$$c_nS(f) = \{fM_1^{\alpha_1}, fM_2^{\alpha_2}, \ldots, fM_n^{\alpha_n}\}$$

For natural numbers $n$ and $m$, we define the languages *sat* $CNF(n, m)$ and $sC(n, m)$:

*sat* $CNF(n, m) = \{\langle f_n \rangle \,/\, f$ is in $CNF$ with $m$ clauses and $f$ is a satisfiable function$\}$,

$sC(n, m) = \{\langle M_1^{\alpha} \,\square\, M_1^{1-\alpha} \star \ldots \star M_i^{\alpha} \,\square\, M_i^{1-\alpha} \star \ldots \star M_n^{\alpha} \,\square\, M_n^{1-\alpha} \rangle \in d(n, m) /$ there are $M_1^{\alpha_1}, \ldots, M_n^{\alpha_n}$, such that $\bigcup_{i=1}^{n}\{M_i^{\alpha_i}\} = \bigcup_{i=1}^{n}(\{M_i^{\alpha}\} \cup \{M_i^{1-\alpha}\})\}$.

*Theorem* 2.7. For any natural number $n$ and $m$,

$$f_n \in sat\,CNF(n, m) \quad \text{if and only if} \quad T_1(f_n) \in sC(n, m).$$

*Proof.* Let $f_n \in satCNF(n, m)$ for some natural $n$ and $m$, and $f$ is the function corresponding to the string $f_n$.

According to the definition, this means that $f$ is represented in $CNF$ with $m$ clauses and $f$ is a satisfiable function. That is, there are some $\sigma_1, \ldots, \sigma_n$, $(\sigma_j \in \{0,1\})$ such, that

$$f(\sigma_1, \ldots, \sigma_n) = 1.$$

We show, that the set $\{fM_1^{\sigma_1}, fM_2^{\sigma_2}, \ldots, fM_n^{\sigma_n}\}$ is a special covering for the set $S(f)$.

By Proposition 2.3 and Lemma 2.6, it suffices to prove that

$$\cup_{i=1}^{n} fM_i^{\sigma_i} = S(f).$$

Let's show that for each clause $e_j$ there is a subset $fM_i^{\sigma_i} \subseteq S(f)$, which includes $e_j$.

Suppose there is a clause $e_j$ not included in $fM_i^{\sigma_i}$, for any $i \in \{1, \ldots, n\}$. This means that none of the literals $x_1^{\sigma_1}, x_2^{\sigma_2}, \ldots, x_n^{\sigma_n}$ is found in the clause $e_j$. It means $e_j$ is a disjunction of some literals of the form $x_i^{1-\sigma_i}$ for $i \in \{1, \ldots, n\}$. Since $\sigma_i^{1-\sigma_i} = 0$ for any $i \in \{1, \ldots, n\}$, then for given assignment of variables, $e_j = 0$. And this is contrary to the fact that

$$f(\sigma_1, \sigma_2, \ldots, \sigma_n) = 1.$$

Hence, each clause enters into some subset $fM_i^{\sigma_i}$ included in the set

$$\{fM_1^{\sigma_1}, fM_2^{\sigma_2}, \ldots, fM_n^{\sigma_n}\}.$$

So, this set is a special covering of $S(f)$. That is,

$$\cup_{i=1}^{n} M_i^{\sigma_i} = \cup_{i=1}^{n}(\{M_i^{\alpha}\} \cup \{M_i^{1-\alpha}\}).$$

Therefore

$$T_1(f_n) = \langle fM_1^{\alpha} \circ fM_1^{1-\alpha} \star fM_2^{\alpha} \circ fM_2^{1-\alpha} \star \ldots \star fM_n^{\alpha} \circ fM_n^{1-\alpha} \rangle \in sC(n, m).$$

Suppose now that for some $n$ and $m$, $T_1(f_n) \in sC(n, m)$.

It means $T_1(f_n) \in d(n, m)$ and there are $\alpha_1, \alpha_2, \ldots, \alpha_n$ such, that the set

$$\{fM_1^{\alpha_1}, fM_2^{\alpha_2}, \ldots, fM_n^{\alpha_n}\}$$

is a special covering for the set $S(f)$.

According to definition, the subset $fM_i^{\alpha_i}$ contains clauses that contain the literal $x_i^{\alpha_i}$.

Therefore, if $x_i^{\alpha_i} = 1$, then the value of all clauses in $fM_i^{\alpha_i}$, is equal to 1:

$$\forall i\ \forall j\ [(\{i, j\} \subseteq \{1, \ldots, n\})\ \&\ (x_i^{\alpha_i} = 1)\ \&\ (e_j \in fM_i^{\alpha_i})] \Rightarrow (e_j = 1).$$

Obviously, if $\sigma_1 = \alpha_1, \sigma_2 = \alpha_2, \ldots, \sigma_n = \alpha_n$, then $f(\sigma_1, \sigma_2, \ldots, \sigma_n) = 1$.

Comparing this with the results of Lemma 2.6, that is:

$$T_1(f_n) \in d(n, m) \quad \text{if and only if} \quad f_n \in CNF(n, m),$$

we obtain that $f_n \in sat\,CNF(n, m)$. $\nabla$

*Remark* 2.8. It is not difficult to simulate a Turing machine that recognizes the strings of language $CNF(n, m)$ in a polynomial time with respect to the length of string.

The general workflow of such a $TM$ is as follows:

Similar to $T_1$, it runs over all symbols of input string $w \in \Sigma^*$ and compares the contents of those string with conditions in definition of the language $CNF(n,m)$.

That is, receiving the string $w$, the $TM$ does:

i) checks if the string $w$ consists of clauses separated by symbol '⋆' and arranged in a specific order,

ii) checks if each clause that occurs in string $w$ satisfies the conditions of the language $CLAUSE(n)$,

iii) checks if the number of literals with pairwise different indexes in clauses included in the string $w$ is equal to $n$.

If any of those conditions is not satisfied, then the $TM$ rejects the string.

It is easy to prove that $TM$ performs the procedures described in points i) - iii), in no more than $c \times |w|^2$ actions for some constant $c$, where $|w|$ is the length of the string $w$.

So, in a polynomial time, such a $TM$ recognizes the string of language $CNF(n,m)$.

Let's for any natural numbers $n$ and $m$, define the function $R_1 : \Sigma^* \to \Sigma^*$ as follows:

$$R_1(w) = \begin{cases} T_1(w), & if \;\; w \in CNF(n,m) \\ \langle \varepsilon \, \square \, \varepsilon \rangle, & if \;\; w \notin CNF(n,m) \end{cases}.$$

It is evident, that $R_1$ is a polynomial time computable function with computable time not exceeded the number $c \times |w|^2$ for certain constant $c$, where $|w|$ is the length of string $w$.

Also, it is evident, that for any string $w \in \Sigma^*$ and for natural numbers $n$ and $m$,

$$w \in CNF(n,m) \text{ if and only if } R_1(w) \in d(n,m).$$

Recall, that $\langle \varepsilon \, \square \, \varepsilon \rangle \in \Sigma^*$ and $\langle \varepsilon \, \square \, \varepsilon \rangle \notin d(n,m)$.

*Theorem* 2.9. For any natural numbers $n$ and $m$,

$$sat\, CNF(n,m) \leq_p sC(n,m).$$

*Proof*: It is obvious that

$$sat\, CNF(n,m) \subseteq CNF(n,m) \text{ and } sC(n,m) \subseteq d(n,m)$$

According to Lemma 2.4, there is a Turing's machine $T_1$, which for any string

$$f_n \in CNF(n,m),$$

forms the string $T_1(f_n)$ in polynomial time with respect to the length of string $\langle f_n \rangle$.

According to Theorem 2.7 and Remark 2.8, for any string $w$, and for any natural numbers $n$ and $m$,

$$f_n \in sat\, CNF(n,m) \text{ if and only if } R_1(f_n) \in sC(n,m).$$

Since $R_1$ is a polynomial time computable function, then the language $sat\, CNF(n,m)$ is polynomially reduced to the language $sC(n,m)$. ∇

Consider now the reducibility of the problem of finding a special covering for a set, to decidability of the sat$CNF$ problem.

Suppose, that $\{e_1, e_2, \ldots, e_m\}$ is a nonempty set of $m$ elements that are composed over the $\Sigma_2$ as mentioned above.

Let $\langle dS \rangle$ is some string of the language $d(n, m)$:

$$\langle dS \rangle = \langle\, M_1^{\alpha} \circ M_1^{1-\alpha} \star \ldots \star M_i^{\alpha} \circ M_i^{1-\alpha} \star \ldots \star M_n^{\alpha} \circ M_n^{1-\alpha} \,\rangle \in d(n, m).$$

We denote by $N(dS)$ the number of symbols forming the string $\langle dS \rangle$. We call the number $N(dS)$ the length of the string $\langle dS \rangle$.

Based on the string $\langle dS \rangle$ we form some string belonging to the language $CNF(n, m)$, as follows:

Let for each element $e_i \in S$, $L(e_i)$ be the set of literals, that are composed as follows:

$$L(e_i) = \{\, x_j^{\alpha_j} \,/\, (\, j \in \{1, \ldots, n\}) \,\&\, (\, \alpha_j \in \{0,1\}\,) \,\&\, (x_j^{\alpha_j} \in \Sigma_1) \,\&\, (\, e_i \in \{M_j^{\alpha_j}\})\}.$$

Let $l(e_i)$ be the clause formed by literals of the set $L(e_i)$ and $c(e_i)$ be the string over the alphabet $\Sigma_1$, which forms by the set $L(e_i)$:

$$c(e_i) = \langle\, x_{i_1}^{\alpha_1}\, x_{i_2}^{\alpha_2} \ldots x_{i_k}^{\alpha_k} \,\rangle, \text{ where } x_{i_j}^{\alpha_j} \in L(e_i).$$

Let's compose the string $\langle\, c(e_1) \star c(e_2) \star \ldots \star c(e_m) \,\rangle$ over the alphabet $\Sigma_1$, which will correspond to the Boolean function in $CNF$ with $m$ clauses denoted by

$$g(dS) = \wedge_{i=1}^{m} l(e_i).$$

It is easy to see, that as a result of formation of all clauses $c(e_i)$, the number of literals in clauses with pairwise different indexes will be equal to $n$. Therefore,

$$\langle\, c(e_1) \star c(e_2) \star \ldots \star c(e_m) \,\rangle \in CNF\,(n, m).$$

*Lemma* 2.10. If for natural numbers $n$ and $m$,

$$\langle dS \rangle = \langle\, M_1^{\alpha} \circ M_1^{1-\alpha} \star \ldots \star M_i^{\alpha} \circ M_i^{1-\alpha} \star \ldots \star M_n^{\alpha} \circ M_n^{1-\alpha} \,\rangle \in d(n, m).$$

then the string

$$\langle c(e_1) \star c(e_2) \star \ldots \star c(e_m) \rangle$$

is forming in a polynomial time with respect to the length of string $\langle dS \rangle$.

*Proof*: The sketch of proof is similar with the proof of Lemma 2.4.

We describe the general features of the workflow of a particular Turing machine denoted by $T_2$, with the alphabet $\Sigma$, which in polynomial time outputs the string

$$\langle\, c(e_1) \star c(e_2) \star \ldots \star c(e_m) \,\rangle,$$

when inputs the string

$$\langle\, M_1^{\alpha} \circ M_1^{1-\alpha} \star \ldots \star M_i^{\alpha} \circ M_i^{1-\alpha} \star \ldots \star M_n^{\alpha} \circ M_n^{1-\alpha} \,\rangle.$$

We consider a Turing machine $T_2$, with input tape, output tape and two working tapes.

The general principles of $T_2$'s operation are as follows:

To form the clause $c(e_i)$, it runs over the symbols of input data and with finding the element $e_i$ included in $M_j^{\alpha_j}$, adds the literal $x_j^{\alpha_j} \in \Sigma$ to the literals forming the clause $c(e_i)$ on the output tape.

At the first stage of work, $T_2$ records pairs $(j, \alpha_j)$ of indexes of all subsets $M_j^{\alpha_j}$ on the first working tape in the same order in which the subsets are on the input tape.

The second stage of work consists of forming the clauses $c(e_i)$ for $1 \leq i \leq m$.

In order to generate a clause matching the element $e_i$, $T_2$ uses the second working tape, adding $e_i$ on it and watching $e_i$ during the procedure of formation $c(e_i)$.

After recording the pairs $(j, \alpha_j)$, $T_2$ starts work by recording $e_1$ on the second working tape.

To determine each step of $T_2$ it is enough to consider the configuration of three current scanned records on different tapes, which we denote by $(R_1, R_2, R_3,)$, where

$R_1$ is the scanned symbol on input tape,

$R_2$ is the scanned pair of indexes on the first working tape of considering subset,

$R_3$ is the scanned element on the second working tape, that is, the element $e_i$ if the clause $c(e_i)$ is forming.

The first configuration that will be considered to start working over the input data, is

$$(e_k, (1, \alpha), e_1),$$

where $e_k$ is the first symbol of input data.

To form $c(e_i)$, the Turing's machine $T_2$ performs actions depending on configuration: with configurations $(\varepsilon, (j, \alpha_j), e_i)$ or $(e_k, (j, \alpha_j), e_i)$, where $k \neq i$,

- scans the next symbol on input tape.

with configuration $(e_i, (j, \alpha_j), e_i)$,

- adds the literal $x_j^{\alpha_j}$ on output tape,
- scans the next symbol on input tape.

with configuration (▫, $(j, \alpha)$, $e_i$),

- scans the next symbol on input tape,
- scans the pair $(j, 1\text{-}\alpha)$ on the first working tape.

with configuration (⋆, $(j, 1\text{-}\alpha)$, $e_i$)

- scans the next symbol on input tape,
- scans the pair $(j+1, \alpha)$ on the first working tape.

with configuration (*end of input data*, $(j, 1\text{-}\alpha)$, $e_i$) and $i < m$,

- adds ⋆ to the output tape,
- adds the symbol $e_{i+1}$ to the second working tape and scans it,
- scans the first symbol of input tape,
- scans the pair $(1, \alpha)$ on the first tape,
- starts to run over the input data again, to detect the element $e_{i+1}$.

with configuration (*end of input data*, $(n, 1\text{-}\alpha)$, $e_m$), $T_2$ stops.

Obviously, on the output tape we will obtain the string

$$\langle c(e_1) \star c(e_2) \star \ldots \star c(e_m) \rangle.$$

It is easy to see that the number of actions performed by $T_2$ to complete the described procedure is polynomial with respect to the number $N(dS)$. The number of mentioned actions does not exceed the number $c \times (N(dS))^2$ for certain constant $c$. ∇

Let's denote by $T_2(dS)$ the string formed as a result of $T_2$'s work on input $\langle dS \rangle$:

$$T_2(dS) = \langle\ c(e_1) \star c(e_2) \star \ldots \star c(e_m)\ \rangle.$$

Obviously, $T_2$ forms the string $T_2(dS)$ based on the string $\langle dS \rangle$ for any special decomposition $dS$ of the set $\{ e_1, e_2, \ldots, e_m \}$.

*Theorem* 2.11. For any natural numbers $n$ and $m$,

$$\langle dS \rangle \in sC(n, m) \quad \text{if and only if} \quad T_2(dS) \in sat\,CNF(n, m).$$

*Proof.* Let $\langle dS \rangle \in sC(n, m)$. This means that:

i) the total number of elements with pairwise different indexes in all $M_i^{\alpha_i}$, for $\alpha_i \in \{0, 1\}$ and $i \in \{1, \ldots, n\}$), is equal to $m$. So, each of the elements $e_1, e_2, \ldots, e_m$ is included in some of subsets forming the string $\langle dS \rangle$.

ii) for some $\alpha_1, \alpha_2, \ldots, \alpha_n \in \{0,1\}$, there are subsets

$$M_1^{\alpha_1},\ M_2^{\alpha_1}, \ldots, M_n^{\alpha_n} \text{ such that } \bigcup_{i=1}^{n} M_i^{\alpha_i} = \bigcup_{i=1}^{n}(M_i^{\alpha} \cup M_i^{1-\alpha}).$$

This means that for each element $e_i \in \bigcup_{i=1}^{n}(M_i^{\alpha} \cup M_i^{1-\alpha})$, there exists a subset

$$M_j^{\alpha_j} \in \{M_1^{\alpha_1},\ M_2^{\alpha_1}, \ldots, M_n^{\alpha_n}\} \text{ such, that } e_i \in M_j^{\alpha_j}.$$

On the other hand, for any index $i \in \{1, \ldots, m\}$, if $e_i \in M_j^{\alpha_j}$, then the literal $x_j^{\alpha_j}$ is found in the clause $l(e_i)$. Thus, the number of forming clauses will be $m$, which means that the number of strings corresponding to clauses also will be $m$.

Let's for simplicity denote by $g_n = T_2(dS)$. Also, we denote by $g$ the function corresponding to the string $g_n$.

Since $x_j = \alpha_j$ implies $x_j^{\alpha_j} = 1$, then $\sigma_1 = \alpha_1\ \&\ \sigma_2 = \alpha_2\ \&\ \ldots\ \&\ \sigma_n = \alpha_n$ implies

$$g(\sigma_1, \ldots, \sigma_n) = g(dS)(\sigma_1, \ldots, \sigma_n) = 1. \text{ So,}$$

$$T_2(dS) = \langle\ c(e_1) \star c(e_2) \star \ldots \star c(e_m)\ \rangle \in sat\,CNF(n, m).$$

Now suppose, that $T_2(dS) \in sat\,CNF(n, m)$. That is, $g$ is a satisfiable function, which means that for some $\sigma_1, \ldots, \sigma_n$,

$$g(\sigma_1, \sigma_2, \ldots, \sigma_n) = 1.$$

According to Theorem 2.7, $g_n \in sat\,CNF(n, m)$ if and only if

$$T_1(g_n) = \langle gM_1^{\alpha} \circ gM_1^{\alpha} \star gM_2^{\alpha} \circ gM_2^{\alpha} \star \ldots \star gM_n^{\alpha} \circ gM_n^{\alpha} \rangle \in sC(n, m).$$

Therefore, there are subsets $gM_1^{\sigma_1},\ gM_2^{\sigma_2}, \ldots, gM_n^{\sigma_n}$ such, that

$$\bigcup_{i=1}^{n} gM_i^{\sigma_i} = \bigcup_{i=1}^{n}(gM_i^{\alpha} \cup gM_i^{1-\alpha})$$

This means that for every clause $l(e_i)$, there exists a subset $gM_j^{\sigma_j}$ such that

$$gM_j^{\sigma_j} \in \{gM_1^{\sigma_1}, gM_2^{\sigma_2}, \ldots, gM_n^{\sigma_n}\} \text{ and } l(e_i) \in gM_j^{\sigma_j}.$$

But then $e_i \in M_j^{\sigma_j}$ , because of definition of $gM_j^{\sigma_j}$:

$$gM_j^{\sigma_j} = \{l(e_k) \,/\, l(e_k) \in S(g) \text{ and } l(e_k) \text{ contains } x_j^{\sigma_j}, \ (k \in \{1, \ldots, m\})\}.$$

At the same time the clause $l(e_i)$ contains the literal $x_j^{\sigma_j}$ only if $e_i \in M_j^{\sigma_j}$.

Since each element $e_i$ determines the composition of one clause, and each clause is defined by one element, then it is easy to prove that for any element $e_i$ there exists a subset

$$M_j^{\sigma_j} \in \{M_1^{\sigma_1}, M_2^{\sigma_2}, \ldots, M_n^{\sigma_n}\} \text{ such, that } e_i \in M_j^{\sigma_j}.$$

$$\text{So, } \bigcup_{i=1}^{n} M_i^{\sigma_i} = \bigcup_{i=1}^{n}(M_i^{\alpha} \cup M_i^{1-\alpha}).$$

Therefore, $\langle\, M_1^{\alpha} \,\text{▫}\, M_1^{1-\alpha} \star \ldots \star M_i^{\alpha} \,\text{▫}\, M_i^{1-\alpha} \star \ldots \star M_n^{\alpha} \,\text{▫}\, M_n^{1-\alpha} \,\rangle \in sC(n, m)$. ∇

*Remark* 2.12. It is not difficult to simulate a Turing machine that recognizes the string of language $d(n, m)$ in a polynomial time.

The general workflow of such a $TM$ is as follows:

Similar to $T_2$, it runs over all symbols of input string $w \in \Sigma^*$ and compares the composition of those string with conditions in definition of the language $d(n, m)$.

That is, receiving the string $w$, the $TM$ does:

i). checks if the string $w$ composes by strings $\langle\, M_i^{\alpha} \,\text{▫}\, M_i^{1-\alpha} \,\rangle$ and of symbols '$\star$',

ii). checks if the number of elements with pairwise different indices in the string $w$ is equal to m,

iii). for each $i \in \{1, \ldots, n\}$, checks if $\langle\, M_i^{\alpha} \,\text{▫}\, M_i^{1-\alpha} \rangle \in PAIR(m)$.

If any of those conditions is not satisfied, then the $TM$ rejects the string.

To perform the procedure of point i) and point ii), it is enough for $TM$ to run over the string, once.

To perform the procedure of point iii) for certain $i$, $TM$ needs to check if takes place

$$(M_i^{\alpha} \in SAB^{\varepsilon}(m)) \ \& \ (M_i^{1-\alpha} \in SAB^{\varepsilon}(m)) \ \& \ (\{M_i^{\alpha}\} \cap \{M_i^{1-\alpha}\} = \emptyset) \ \& \ (M_i^{\alpha} \neq \varepsilon \text{ or } M_i^{1-\alpha} \neq \varepsilon)\}.$$

If we denote by $m_i$ the total number of elements included in subsets $\{M_i^{\alpha}\}$ and $\{M_i^{1-\alpha}\}$ then it is easy to see, that for each $i \in \{1, \ldots, n\}$, the number of actions for checking the relation

$$\langle\, M_i^{\alpha} \,\text{▫}\, M_i^{1-\alpha} \,\rangle \in PAIR(m)$$

does not exceed $c \times m_i^2$, for some constant $c$. It is important to note, that for any $i$, $m_i \leq m$.

Then, it is evident that $TM$ performs the procedure described in point iii) in no more than $c \times |w|^2$ actions for some constant $c$, where $|w|$ is the length of the string $w$.

So, in a polynomial time, such a $TM$ recognizes the string of language $d(n, m)$.

Let's for natural numbers $n$ and $m$, define the function $R_2 : \Sigma^* \to \Sigma^*$ as follows:

$$R_2(w) = \begin{cases} T_2(w), & if \ \ w \ \in d(n,m) \\ \langle \varepsilon \circ \varepsilon \rangle, & if \ \ w \ \notin d(n,m) \end{cases}.$$

It is evident, that $R_2$ is a polynomial time computable function with computable time not exceeded the number $c \times |w|^2$ for certain constant $c$, where $|w|$ is the length of string $w$.

It is also obvious that for any natural numbers $n$ and $m$, and for any string $w \in \Sigma^*$,

$$w \in d(n,m) \text{ if and only if } R_2(w) \in CNF(n,m).$$

Recall, that $\langle \varepsilon \circ \varepsilon \rangle \in \Sigma^*$ and $\langle \varepsilon \circ \varepsilon \rangle \notin d(n, m)$.

*Theorem* 2.13. For any natural numbers $n$ and $m$,

$$sC(n, m) \leq_p sat\,CNF(n, m)$$

*Proof.* It is obvious that

$$sC(n, m) \subseteq d(n,m) \text{ and } sat\,CNF(n,m) \subseteq CNF(n,m).$$

According to Lemma 2.10, for any natural numbers $n$ and $m$, there is a Turing's machine $T_2$, which for any string

$$\langle dS \rangle = \langle\, M_1^{\alpha} \circ M_1^{1-\alpha} \star \ \ldots \star M_i^{\alpha} \circ M_i^{1-\alpha} \star \ldots \star M_n^{\alpha} \circ M_n^{1-\alpha} \,\rangle \in d(n,m)$$

forms the string

$$T_2(dS) = \langle\, c(e_1) \star c(e_2) \star \ldots \star c(e_m) \,\rangle$$

in polynomial time with respect to the length of string $\langle dS \rangle$.

According to the Theorem 2.11 and Remark 2.12, for any string $w$, and for any natural numbers $n$ and $m$,

$$\langle dS \rangle \in sC(n, m) \text{ if and only if } R_2(dS) \in sat\,CNF(n, m).$$

Therefore

$$sC(n, m) \leq_p sat\,CNF(n, m).$$

Since $R$ is a function computable in polynomial time, then the language $sC(n, m)$ is polynomially reduced to the language $sat\,CNF(n, m)$. ∇

*Theorem.* 2.14. For any natural numbers $n$ and $m$, the languages

$$sat\,CNF(n, m) \text{ and } sC(n, m)$$

are polynomially equivalent.

*Proof.* Follows of Theorems 2.9 and 2.13.

*Corollary* 2.14.1 The problem of existence of a special covering for a set under the special decomposition of this set is an $NP$ - complete problem.

*Proof.* Follows from the Theorem 2.14. ∇